\documentstyle[psfig]{aa}
\newcommand{\Msun}{M$_{\odot}$}
\begin{document}
\thesaurus{08   
                (11.09.1 NGC 5907;
                11.08.1;
                11.09.2;
                11.01.1;
                11.11.1;
                11.16.1)}
\title {NGC\,5907 revisited: a stellar halo formed by cannibalism? \thanks{
Based on data obtained with the CFHT in Hawaii}}

\author{J. Lequeux\inst{1}\and
        F. Combes\inst{1}\and
        M. Dantel-Fort\inst{1}\and
        J.-C. Cuillandre\inst{2}\and
        B. Fort\inst{1}\inst{3}\and
        Y. Mellier\inst{1}\inst{3}}
\offprints{lequeux@mesioa.obspm.fr}
\institute{DEMIRM, Observatoire de Paris, 61 Avenue de l'Observatoire, F-75014
Paris, France
\and
Canada-France-Hawaii Telescope Corporation, P.O. Box 1597, Kamuela, HI 96743,
USA
\and
Institut d'Astrophysique, 98 bis Bd. Arago, 75014 Paris, France}
\date{Received ......; accepted ......}
\maketitle
\begin{abstract}
We report on further observations of the luminous halo of NGC\,5907. New
V, I and B deep photometry confirms the existence of an extended stellar
halo redder than the disk. Our data are consistent with a faint halo,
or very thick disk, composed of a metal-rich old stellar population.
We propose that it could be the remnant of a merged
small elliptical, and we support our hypothesis with N-body simulations.
\end{abstract}
\section{Introduction}
Sackett et al. (\cite{Sackett}, hereafter SMHB) have found that the edge--on
spiral galaxy NGC\,5907 is surrounded by a faint halo in the optical R band.
The existence of this halo has been confirmed by several authors at
other wavelengths. Lequeux et al. (\cite{PaperI}, hereafter
Paper I), have observed this halo in two other bands, V and I, and suggest that
it becomes redder at increasing distances from the plane of the galaxy.
Rudy et al. (\cite{Rudy}) and James \& Casali (\cite{James})
have also detected the halo in J and K; the combination of
their brightnesses with those of SMHB and of
Paper I gives results difficult to understand. The halo of NGC\,5907 is
clearly stellar in origin but its nature is still unknown. Its relatively red
V-I color implies either a normal, metal-rich, old stellar population, or if the
stars are metal-poor an initial mass function favoring extremely low masses
(see discussion in Paper I).
The observations being very difficult because of the faintness of the halo, we
decided to perform new observations, described in Sect. 2 and 3. To interpret
them, Sect. 4 proposes a scenario in which NGC 5907 has
encountered and cannibalized a small elliptical galaxy about 2 Gyr ago. 

\section{Observations}

All the observations have been obtained with the Canadian-France-Hawaii
Telescope. The V and I
observations were made in a single photometric night and used the UH 8k
mosaic of CCD at the prime focus.
This camera being insensitive in the blue, the B observations were performed
during another night with the MOS
instrument at the Cassegrain focus, equiped with the STIS-2 2048$\times$2048
pixel CCD camera.

The UH 8k mosaic is made of eight 2048x4096 pixel CCDs arranged to form a
square. The scale is 0.206 arc second per pixel of 15 $\mu$m, but the pixels
were binned so that the final pixels are of 0.412 arc second. We used 
a position--switch observing mode in order to limit the effects of the changes
in the sky background. The image of the galaxy was centered parallel to the 
longer side of one of the CCDs during half of the exposures, and moved
to the parallel CCD during the other half. Consequently, only two of the eight
CCDs were used but the field covered by each of them ($14\arcmin\times7\arcmin$)
was sufficient for our purpose. When the galaxy is imaged on one CCD, the other
CCD is used for defining the flat field used to reduce the images
obtained in the alternative situation. We obtained in each of the V and
I$_{Cousins}$
band 9 frames of 480 s each at one position alternating with 9 similar frames at
the other position. The frames at a given position were taken with slight
shifts with respect to each other in order to allow elimination of the
spurious events through a median stacking
of the images. The flat fields obtained in this way were of excellent quality
and we noticed a significant reduction of the large--scale background
irregularities in the final images with respect to the observations
of Paper I. This can be attributed
to a partial cancellation of the diffuse light by our position--switch
technique.
The data were calibrated using reference stars in SA 110 (Landolt
\cite{Landolt}).
The photometric error is of the order of 0.03 magnitude in each band. The
sky levels measured on our frame were 21.43 and 19.22 mag.arcsec$^{-2}$ in
V and I respectively, in excellent agreement with the mean values for
photometric nights at Mauna Kea.

The rest of the data processing was identical to that of
Paper I. After masking all discrete objects seen in the frame, the surface
brightness was averaged in rectangles parallel to the major axis of NGC\,5907.
The length of these rectangles is 3.84$\arcmin$ (as for fig. 3 of Paper I),
corresponding to 12.3 kpc at the adopted distance of 11 Mpc. 

The B observations have been made in the classical way with the MOS focal
reducer at the Cassegrain focus of the CFHT. Baffles have been inserted
in this device to reduce the scattered light. The 21 $\mu$m pixels
of the CCD correspond to 0.44$\arcsec$. 8 exposures of 900 seconds
each have been obtained, with slight shifts from an exposure to the next
(as above). 
The calibrations were also made on SA 110. The rest of the reductions was done
as for the V and I images. The final surface brightness was averaged in
the same rectangles as for V and I.

\begin{figure}
\psfig{figure=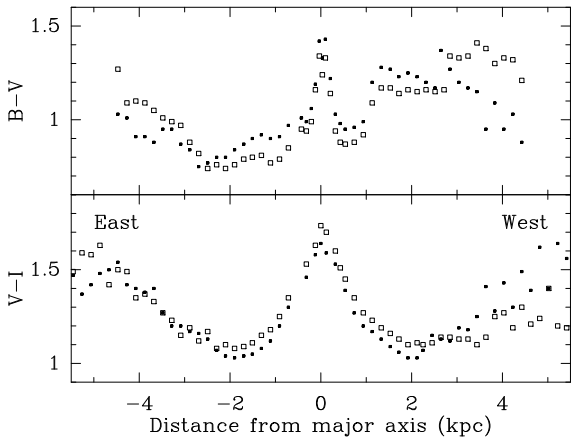,bbllx=0cm,bblly=75mm,bburx=18cm,bbury=205mm,width=8cm}  
\caption{The V-I$_{Cousins}$ (bottom) and the B-V (top) profiles perpendicular
to the major axis of NGC\,5907. The results are averaged over long rectangles
(3.84$\arcmin$) parallel to the major axis and centered on the minor axis of the
galaxy. The two sets of different symbols correspond to position--switch V
and I observations in which the galaxy was located on one or the other of the
two CCDs: see text. Their respective scatter give an idea of the uncertainties
on the results, which are difficult to estimate objectively.}
\end{figure}

\section{Observational results}
Figure 1 presents the V-I and B-V color profiles perpendicular to the plane of
NGC\,5907. We do not show the surface brightness profiles themselves as
they are in excellent agreement with those displayed on fig. 2 of Paper I.
Note that the orientation given in the figure caption of fig. 1 of Paper I
is incorrect: this fig. 1 displays
a mirror image with North to position angle approx. 150$^{\circ}$ and East
to 60$^{\circ}$.
The colors are strongly affected near the galactic plane by the reddening due to
the interstellar dust layer of the disk. The
B--V color to the right of fig. 1 is also redder than on the other side
and might suggest extinction. NGC 5907
is not exactly edge--on. Its inclination is 86$^{\circ}$ according to Rand
(\cite{Rand});
his fig. 7 shows clearly an absorption band shifted to the West of the plane of
the galaxy.  Its strong extinction starts at 1 kpc from the plane,
and a jump in B--V (actually in the B band) is seen at this position on fig. 1.
However the V and I bands are apparently not affected, and the extinction test
on the colors of background galaxies described in Paper I does not suggest much
extinction on this side of the galaxy. The problem is entirely due to the
B-band image and we have no way of checking the corresponding observation,
contrary to the V and I ones. It may come from the imperfect subtraction of a 
very bright star in this region of the galaxy. Whatever the cause the data for 
the right (West) side are not usable. In what follows, we will only consider 
the left (East) side. 

At about 2 kpc from the plane on this side the effects of extinction are 
minimal. We observe there V-I = 1.0 and B-V = 0.8, and a comparison with
SMHB and Rudy et al. (\cite{Rudy}) gives
B-R $\simeq$ 0.8, B-J $\simeq$ 3.2 and B-K $\simeq$ 3.8. These
colors are typical for the disk of a spiral galaxy (de Jong ~\cite{de Jong}) 
although the flux observed in the R band seems 
comparatively too large by roughly 0.6 mag.

At larger distances from the plane the colors become redder.
They reach V-I = 1.4 at the limit of secure V and I observations (5.5 kpc) 
and B-V = 1.0 at 4.4 kpc. These colors are
typical for an old, metal--rich population (see e.g.
Poulain \& Nieto ~\cite{Poulain}). The B-V color confirms the result already
suggested from the V-I color in Paper I. The colors obtained by combining our B,
V and I photometry with the R from SMHB or
with the H and K from Rudy et al. (\cite{Rudy}) give less consistent
results. At 95$\arcsec$ (5.1 kpc) from the plane on the East side one has 
V-R $\simeq$ 2.0, V-J $\simeq$ 4.3 and V-K $\simeq$ 5.0, with of course very 
large errors of perhaps 0.5 mag. All these colors are 
redder by 1--2 mag. than those of elliptical galaxies (respectively
about 0.6, 2.3 and 3.2 mag.: Poulain \& Nieto ~\cite{Poulain}; 
Peletier ~\cite{Peletier93}). The result for V-R is 
particularly troublesome since the R band lies between V and I.
This is probably due to the extreme difficulty of these measurements which are 
very sensitive to the adopted sky background and to scattered light.
In what follows we will only consider our B, V and I photometry. 

Then our conclusions are the same as in Paper I. Either
the halo light at a large distance from the plane of NGC\,5907 is dominated by
an old, metal--rich stellar population, or if the stars are more metal--poor as
expected for halo stars, their mass function should be strongly dominated by
very low mass stars of 0.15 \Msun \, or less. The latter hypothesis looks rather
arbitrary, although it would help to solve the problem of the dark matter in 
NGC\,5907. 
Consequently we prefer the former hypothesis, but now the problem is to
understand how a metal--rich, old population can populate the thick system
for which we observe the corresponding colors. We will
show now that this is possible as the result of capture of a small elliptical
galaxy by a large spiral. 
%

\begin{table}
\begin{flushleft}
\caption[]{Parameters of the simulation}
\scriptsize
\begin{tabular}{lcccc}
\hline
\multicolumn{1}{c}{}                     &
\multicolumn{1}{c}{scale-length}                   &
\multicolumn{1}{c}{total mass }      &
\multicolumn{1}{c}{N}                   \\
\multicolumn{1}{c}{}                     &
\multicolumn{1}{c}{ (kpc)}                   &
\multicolumn{1}{c}{(10$^9$ M$_\odot$) }      &
\multicolumn{1}{c}{(particles)}                   \\
 & & &  \\
NGC 5907 &  &  & \\
Bulge  & 1.  & 13.6 & 6723  \\
Disk & 6.  & 113. & 56023 \\
Halo & 20.  & 181. & 89638  \\
Total &   & 307. & 152384\\
 &  & &  \\
Companion & 6.   & 30.7  & 15238 \\
Grand Total & & 338.  & 167622 \\
 &  & & \\
\hline
\end{tabular}
\end{flushleft}
\end{table}
%
\begin{figure}
\psfig{figure=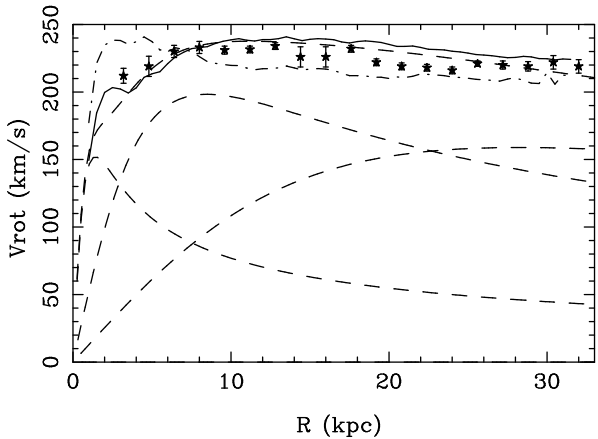,bbllx=2cm,bblly=2cm,bburx=105mm,bbury=135mm,width=8cm,angle=-90}  
\caption[]{Rotation curves from the relaxed initial state of the simulation
(full line), and the final state (dash-dot), 
compared to the observations (stars and error bars from Sancisi \& 
van Albada 1987) and to the analytical curves obtained from the various
components (dashed lines) }
\end{figure}

\begin{figure}
\psfig{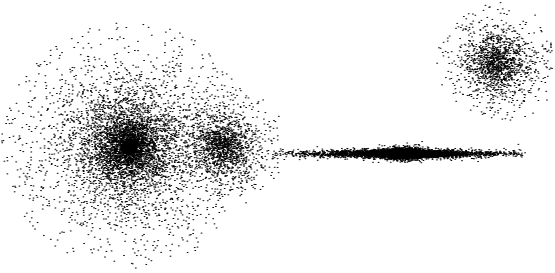}  
\caption[]{Initial configuration; only the visible matter is shown
(face-on and edge-on views). One out of 10 particles is plotted.}
\end{figure}

\section{Merging a small elliptical galaxy with a large spiral}

In order to check this idea, we performed a single N-body simulation of
this capture. The code used was the tree-code from Barnes \& Hut 
(\cite{Barnes1986}, \cite{Barnes1989}).
Table 1 summarizes the initial conditions, i.e. the scale-length
and total mass of each component. All components
were truncated at a radius of 32 kpc. The main galaxy
(simulating the late-type spiral NGC 5907) was composed of a spherical
non-rotating bulge, represented by a Plummer component, an $n=1$
Toomre stellar disk, thickened by a sech$^2$ z-distribution with 
a constant scale height of $z_0 = 500$ pc, and a spherical Plummer halo. 
The program computes the combined potential of all components
for the main galaxy, and solves the Jeans equations to derive the
initial rotational velocity and dispersion. The stellar disk was 
launched with a Toomre parameter $Q$ slightly decreasing with radius, 
from 1.5 at the center to 1 at the end of the disk.
Initially the disk is mildy unstable against spiral and
bar formation. The resulting rotation curve is compatible with the 
observed one (Sancisi \& van Albada \cite{Sancisi}),
as shown in figure 2. The companion is represented
by a spherical Plummer component, without dark halo.
Its mass is one tenth of that of NGC\,5907. It is launched at 32 kpc radius 
from NGC\,5907, with a purely tangential velocity of 200 km s$^{-1}$, 
in a bound, almost circular orbit. The initial configuration is displayed in
fig. 3. Fig. 4 shows some steps of the calculation: 
300 Myr after the beginning of the simulation,
the elliptical galaxy has triggered a spiral structure in NGC\,5907
and starts to be slowed down by dynamical friction. Merging is well
advanced 1 Gyr after the beginning, and we stop the simulation after 1.9 Gyr.
At this time, the spiral galaxy has restored its flat disk structure,
although it is warped and thicker due to the dynamical heating of its stars in
the process. This disk is tilted with respect to the original plane due to the
conservation of total angular momentum. The elliptical companion has
become a fat, hollow and distorted oblate ellipsoid centered on the spiral
galaxy and with a rather similar plane of symmetry. This final stage
is displayed on fig. 5. Fig. 6 shows a cut through the "visible" particles of
both galaxies, perpendicular to the new major axis of the spiral. The elliptical
companion dominates the visible mass
 at heights larger than 4 kpc from the plane, and
there the colors should be close to those of the elliptical galaxy as observed.
We emphasize that the companion chosen for the experiment is 5-10 times
more massive that what is needed to explain the observations; this choice
was necessary to remedy the small dynamical range of N-body simulations,
but is still indicative of the dynamical phenomenon.
Even assuming a constant M/L for the stars of the elliptical, fig. 6 shows 
that the light does not follow the dark matter profile, although the dark
halo density is dominant in this region. In any case, there are
many choices for the r- and z-distribution of the dark matter, given a rotation
curve, and this experiment shows that 
the light can be a poor tracer of the dark matter.

\begin{figure}
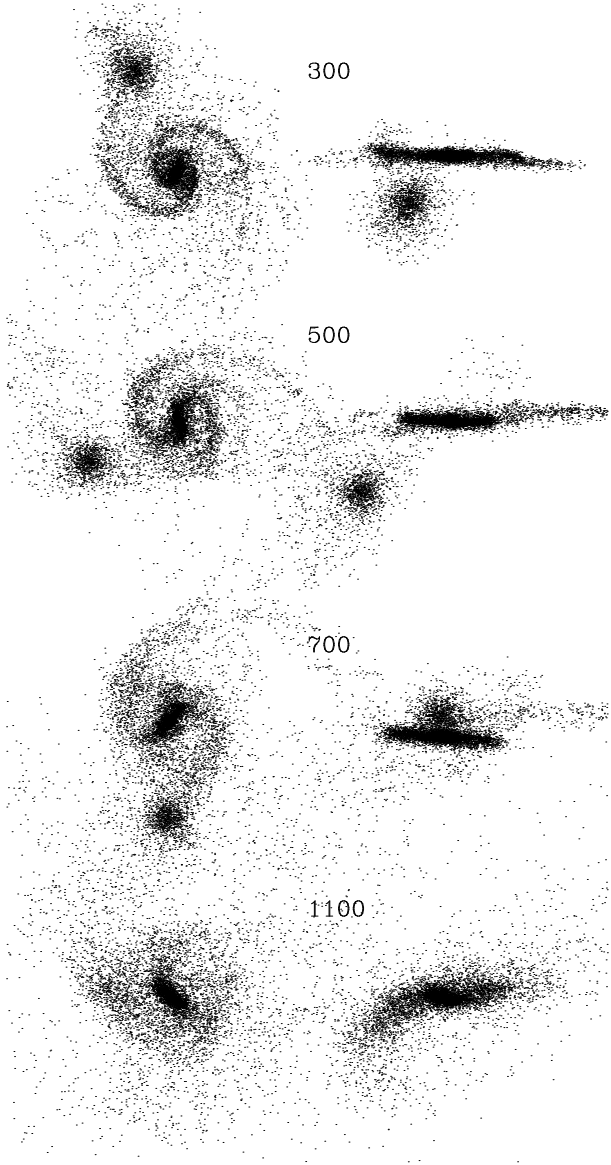

\psfig{figure=Ab161_f4a.ps,bbllx=5cm,bblly=8cm,bburx=17cm,bbury=205mm,width=8cm,angle=-90}  
\psfig{figure=Ab161_f4b.ps,bbllx=5cm,bblly=8cm,bburx=17cm,bbury=205mm,width=8cm,angle=-90}  
\caption[]{ Some steps of the simulations (time in Myrs). }
\end{figure}
\section{Conclusions and perspectives}
In this paper, we reported on new observations made with the CFH telescope which
confirm entirely the conclusions reached in Paper I: the
luminous halo around NGC\, 5907 is redder than the galaxy and has the colors of
a metal--rich old stellar population. 
However discrepancies between the present observations
and others in the literature have to be understood,
which require further observations.
We have shown by a N--body simulation that
this red halo could result from cannibalism of a low--mass elliptical 
(a few 10$^9$ M$_\odot$) after a slow encounter with NGC\, 5907.

\begin{figure}
\psfig{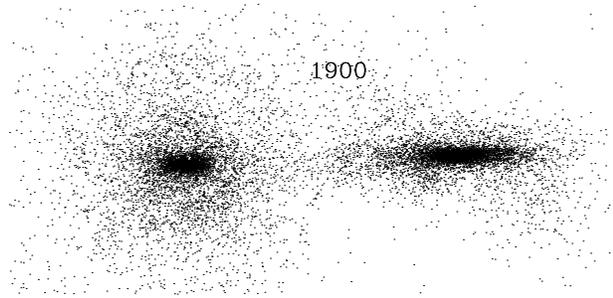}  
\caption[]{  Final plot of the visible stars, when the two galaxies
have merged (here at 1900 Myr). The system is rotated to have its mean angular momentum
vector along the vertical axis.}
\end{figure}

How frequent is this phenomenon? Nearby
spiral galaxies like M 31 often have elliptical companions which 
might eventually merge with them. Our Galaxy might have cannibalized a
satellite which would have produced the thick disk as the result of stellar
dynamical heating, the process evidenced by our simulation (Robin et al.
\cite{Robin}).
To get statistical information on this capture process,
one should conduct a systematic investigation 
of luminous halos, including measurements of their colors, 
around a well--defined sample of edge--on spiral galaxies.

\begin{figure}
\psfig{figure=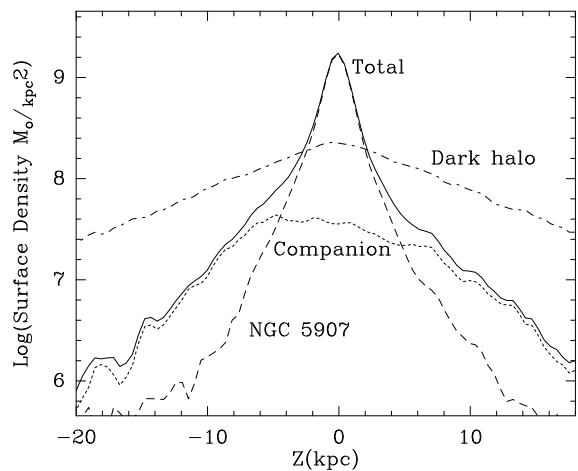,bbllx=0cm,bblly=35mm,bburx=18cm,bbury=18cm,width=8cm}  
\caption[]{ Distribution perpendicular to the plane, 
at a distance of 6kpc from the center,
of the stellar component of NGC 5907 (dash), the companion (dots),
the total visible mass (full line), and dark halo (dot-dash), at t=1900 Myr.}
\end{figure}

\acknowledgements
Simulations have been carried out on the Crays
of IDRIS, the CNRS computing center at Orsay, France.
\end{document}